%% file: main.tex
\documentclass{IOS-Book-Article}     
\usepackage[utf8]{inputenc}
\usepackage[table]{xcolor}
\usepackage{graphicx} 
\graphicspath{ {./Figures/} }
\usepackage[rightcaption]{sidecap}
\usepackage{wrapfig}
\usepackage{multirow}
\usepackage{tabularx}
\usepackage{rotating}
\usepackage{array}
\usepackage{pdfpages}
\usepackage[flushleft]{threeparttable} 
\usepackage{float}
\usepackage{comment}

\newcommand{\eg}{e.g.}

\newcolumntype{C}{>{\raggedright\arraybackslash}X}

\begin{document}
\begin{frontmatter}          
%
\title{Exosoul: ethical profiling in the digital world}
\runningtitle{}

%
\author[univaq]{\fnms{Costanza} \snm{Alfieri}},
\author[univaq]{\fnms{Paola} \snm{Inverardi}}
\author[univaq]{\fnms{Patrizio} \snm{Migliarini}}
\author[univaq]{\fnms{Massimiliano} \snm{Palmiero}}
	\address[univaq]{Universit\`a degli studi dell'Aquila	L'Aquila, Italy\\\{paola.inverardi,  massimiliano.palmiero\}@univaq.it, costanza.alfieri@guest.univaq.it, patrizio.migliarini@graduate.univaq.it}
\begin{abstract}

\input{Abstract}
\end{abstract}


\end{frontmatter}


\section{Introduction}
\label{sec:Introduction}
\input{Introduction}

\section{Experiment}
\label{sec:Experiment}
\input{Experiment}


\section{Discussion}
\label{sec:Discussion}
\input{Discussion}

\section{Conclusions and Future Work}
\label{sec:Conclusions and future work}

\input{Conclusions_and_Future_Work}


\clearpage

\bibliographystyle{plain}
\bibliography{main}


\end{document}

%% file: Abstract.tex
The development and the spread of increasingly autonomous digital technologies in our society pose new ethical challenges beyond data protection and privacy violation. Users are unprotected in their interactions with digital technologies and at the same time autonomous systems are free to occupy the space of decisions that is prerogative of each human being. In this context the multidisciplinary project Exosoul aims at developing a personalized software exoskeleton which mediates actions in the digital world according to the moral preferences of the user. The exoskeleton relies on the ethical profiling of a user, similar in purpose to the privacy profiling proposed in the literature, but aiming at reflecting and predicting general moral preferences. Our approach is hybrid, first based on the identification of profiles in a top-down manner, and then on the refinement of profiles by a personalized data-driven approach. In this work we report our initial experiment on building such top-down profiles. We consider the correlations between ethics positions (idealism and relativism) personality traits (honesty/humility, conscientiousness, Machiavellianism and narcissism) and worldview (normativism), and then we use a clustering approach to create ethical profiles predictive of user's digital behaviors concerning privacy violation, copy-right infringements, caution and protection. Data were collected by administering a questionnaire to 317 young individuals. In the paper we discuss two clustering solutions, one data-driven and one model-driven, in terms of validity and predictive power of digital behavior.

%% file: Keywords.tex
Autonomous Systems; Ethics; User Profiling; Personality; Digital Behaviors


%% file: Introduction.tex

\begin{figure}[!htp]
    \centering
    \includegraphics[width=1\textwidth]{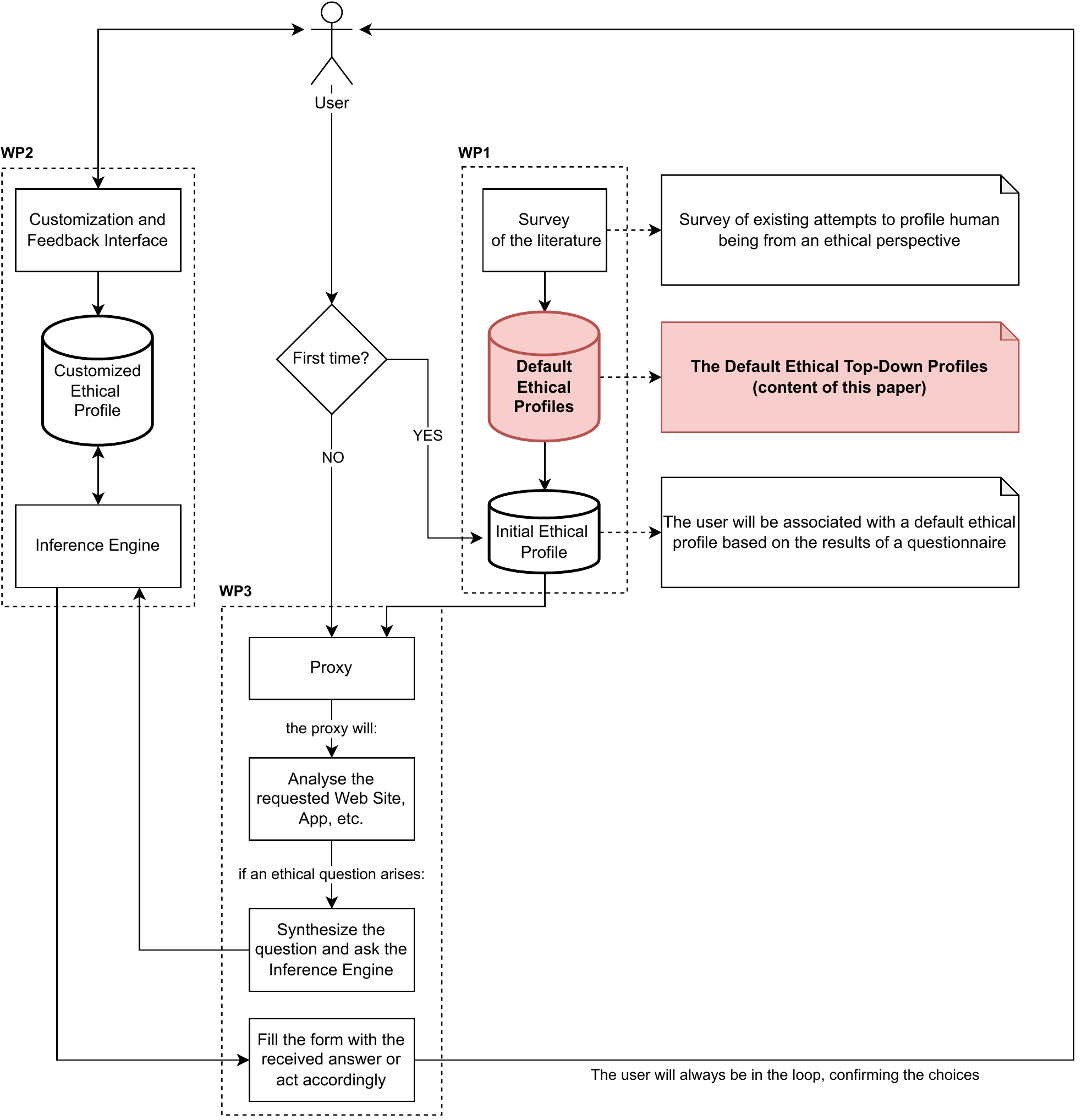}
    \caption{EXOSOUL Ethical Engine Scheme with highlight on the work package related to this research.}
    \label{fig:ethical_engine}
\end{figure}

There is a general and growing consensus that the diffusion of autonomous digital technology may harm the values our societies are based on. Europe is at the forefront of the elaboration on these issues especially from a regulatory point of view. First concerning privacy with the GDPR \cite{GDPR} and currently with the AI act \cite{AIact}. Regulation is important and represents the level of awareness that the society as a whole has matured regarding the potential misuse of digital technologies. However it is not sufficient and cannot cover all the space of potential misuses concerning the risk which attains the core of the fundamental rights of the citizens. Privacy concerns are insufficient: ethics and the human dignity are at stake \cite{InverardiACM,InverardiDIGHUM}. As a matter of fact, individuals are unprotected and powerless in their interaction with the digital world. In a digital society where the relationship between citizens and machines is uneven, moral values like individuality and responsibility are at risk. Despite the ideal of a human centric AI and the recommendations to empower the users, the power and the burden to preserve the users’ rights still remain in the hands of the (autonomous-) systems producers. 
The EXOSOUL \cite{Exosoul} project aims to empower humans with an automatically generated exoskeleton, i.e. a software shield that protects them and their personal data through the mediation of all interactions with the digital world that would result in unacceptable or morally wrong behaviors according to their ethical and privacy preferences.
The exoskeleton relies on the ethical profiling of a user, similar in purpose to
the privacy profiling proposed in the literature \cite{schairer2019disposition, dupree2018case}, but aiming at reflecting general moral preferences and predicting user's digital behaviors accordingly like proposed in \cite{MobileSoft20}. More precisely, EXOSOUL is based on the notion of digital ethics \cite{floridi2018soft} and its separation in soft ethics to reflect user's ethics and hard ethics to define the ethical values a digital system shall comply with.  Our approach is hybrid, first based on the identification of profiles in a top-down manner, and then on the refinement of profiles by a personalized data-driven approach see Fig.\ref{fig:ethical_engine}. In the following we report our initial experiment on building such top-down
profiles.

%% file: Experiment.tex
\subsection{Theoretical Background}

In order to define users' ethical profiles we considered the Ethics Position Theory (EPT)\cite{forsyth1980taxonomy,forsyth2008east, o2021individual}, which suggests individuals' differences in moral judgments. This theory posits that the individuals' personal 
morality contains unique and idiosyncratic elements that depend on the experience related to moral issues. These elements are sustained by two dimensions: idealism and relativism. The former reflects absolute moral principles, mostly oriented to truth, benevolence and avoiding harming others. The latter relies on the careful evaluations of the situations, contexts and consequences, and also reflects the idea that harm is sometimes necessary in order to produce good. These two dimensions are based on the deontological and teleological models, respectively. The deontological perspective defines the rightness and wrongness of possible courses of action, by comparing them with pre-determined norms representing personal values, whereas the teleological perspective considers the perceived consequences, the probability and the desirability of each course of action for various stakeholder groups, and the importance of each group. The EPT does not assume that individuals are only rule-oriented (idealism) or consequence-oriented (relativism), but rather assumes that individuals can range from high to low in their idealism and relativism ideologies. This leads to identify four moral philosophies: situationism, subjectivism, absolutism, and exceptionism (see Table \ref{tab:EPT}, \cite{o2021individual}).

\begin{table}[!htp]
\tiny
 \centering
\begin{tabular}{|p{2 cm}|p{4.5 cm}|p{4.5 cm}|}

\hline
 & \textbf{Low Relativism} & \textbf{High Relativism} \\
 \hline
\raggedright \textbf{Low Idealism} & \textbf{Exceptionists}: Conventionalists who tolerate
exceptions to moral standards when benefits offset
potential harmful consequences. & \textbf{Subjectivists}: Realists who do not endorse moral
standards that define right and wrong or the
avoidance of harmful consequences \\
\hline
\raggedright \textbf{High Idealism} & \textbf{Absolutists}: Principled idealists who endorse both reliance on moral standards and striving to
minimize harm done to others &\textbf{ Situationists}: Idealistic contextualists who value minimizing harm rather than reliance on moral
standards that define right and wrong \\

\hline

\end{tabular}
\caption{The four moral types identified in EPT.}
    \label{tab:EPT}
\end{table}

The EPT has been related to personality, which includes dispositions, temperaments, characters, attitudes, values, and so forth. Personality is “the dynamic organization within the individual of those psycho-physical systems that determine his unique adjustments to the environment” (p. 48)\cite{allport1937personality}, by creating "....the person's characteristic patterns of behavior, thoughts, and feelings" (p. 48)\cite{allport1961pattern}. Although the personality attributes are independent of each other, some can align to form profiles of shared similarities. This means that individuals’ ethics positions are systematically related, amongst others, to personality\cite{forsyth2019making}. The most consolidated approach to personality assumes that a small number of enduring dispositions or traits provides the basis for differences and similarities among people. In this vein, the Five Factor Model (FFM)\cite{costa1992four, costa2013personality} is the most useful framework, given that the 5 personality traits identified have emerged with great regularity in different studies: neuroticism, extraversion, conscientiousness, agreeableness, and openness. Afterward, the Hexaco theory of personality identified a sixth factor: honesty/humility trait\cite{lee2004psychometric, lee2013h}. In addition, the construct of Dark Triad personality was also proposed, based on the idea that the human nature encompasses also a constellation of sub-clinical and malevolent traits\cite{paulhus2002dark}, which reflect different facets of antisocial and adverse personality\cite{petrides2011trait}: psychopathy, Machiavellianism, and narcissism. 

In particular, honesty/humility and conscientiousness traits were found to be mostly associated positively with morality and integrity\cite{cohen2013agreement, perugini2009implicit}, and highly resistant to moral disengagement\cite{cohen2014morala, ogunfowora2021meta, detert2008moral, moore2012employees}. Basically, these two personality traits can give insights into what a moral character is\cite{cohen2014moral}.

Specifically, the honesty/humility trait reflects sincerity, fairness, greed, modesty, and no manipulation, no interested in richness, luxuries, and elevated social status. This trait is characterized by "the tendency to be fair and genuine in dealing with others, in the sense of cooperation with others even when one might exploit others without suffering retaliation” (p. 156)\cite{ashton2007empirical}. It was found related positively to individualizing values (e.g., sensitivity to harm and fairness)\cite{zeigler2015spitefulness}. This leads people with high honesty/humility to conform to idealism rather than to relativism. Indeed, previous studies showed that honesty-humility is correlated positively to idealism and negatively to relativism \cite{cohen2014morala, ogunfowora2021meta}.

Conscientiousness reflects the tendency of being persistent in the pursuit of goals/tasks, and relies on orderliness, meeting of obligations, self-confidence, self-regulation and self-discipline, integrity, fairness, and inhibition of impulses that go against moral obligations. Conscientious people can determine their ethical codes, are less social and tend to sacrifice social ties for achieving goals\cite{mccrae1987validation}. Conscientiousness was related to the exceptionist moral philosophy, being not correlated either to idealism or relativism\cite{khan2016time}. However, other studies showed that this trait was either positively correlated to relativism\cite{forsyth2019making}, or positively to idealism but negatively to relativism\cite{cohen2014morala, ogunfowora2021meta}. 

As concerns the Dark Triad construct, the three malevolent personality traits were found related to moral disengagement and \textit{schadenfreunde} (feeling of pleasure at another’s suffering)\cite{erzi2020dark}. Notably, Machiavellianism and psychopathy are highly correlated, especially in relation to the Big Five personality traits. Thus, these two personality traits are seen as a single psychopathic entity\cite{o2015meta}, whereas narcissism is considered the lightest dimension of the Dark Triad\cite{furnham2013dark}. Besides, these two traits are more predictive of digital behaviors. Machiavellianism and narcissism were found positively related to self-promotion on Facebook\cite{rosenberg2011online}, to the number of personal information disclosed online\cite{sanecka2017dark, pureesmali2017dark}, and to the tendency to self-disclose in computer-based communication\cite{sanecka2017dark}. Machiavellianism and vulnerable narcissism predicted also problematic social media use\cite{kircaburun2019analyzing} and less congruence between the true self and the Instagram self\cite{geary2021insta}.

In details, Machiavellianism reflects actions oriented to pragmatism, entails a lack of affect in interpersonal relationships, and moral standards. High Machiavellianism is associated with manipulative and cynical tactics, the basic attitude being 'the ends justify the means'\cite{christie2013studies, collison2018development}. Machiavellianism is characterized by an utilitarian rather than a moral view when interacting with others\cite{christie2013studies}. Different studies showed that Machiavellianism correlates negatively to idealism but positively to relativism\cite{cohen2014morala, leary1986ethical, mchoskey1999machiavellianism}. Following Leary et al.\cite{leary1986ethical}, Machiavellianism is basically associated with the subjectivism moral philosophy. 

Narcissism is supported by an inflated opinion of self, feelings of entitlement, superiority, need for admiration\cite{emmons1987narcissism}, egoism and arrogance\cite{forsyth2012meta}. It is also characterized by moral disengagement\cite{ogunfowora2021meta}, unethical attitudes and questionable behaviors aimed at pointing out high achievements, which in turn reiterate admiration, power, and ego superiority\cite{goldman2009destructive, penney2002narcissism}. Narcissism was found to correlate positively to idealism but not to relativism\cite{waldman2017neurological}, or to be higher in relativistic egoists cluster than idealistic altruists cluster\cite{chudzicka2013ethical}. Yet, the narcissistic gratification dimensions correlated to idealism but not to relativism\cite{sommerfeld2010subjective}. These contradictory results support the idea that narcissists can behave morally 'right' to increase their ego and get higher levels of narcissistic supplies from helping others, even though they are morally disengaged.

Notably, beside personality, individuals’ ethics positions can be also related to worldviews. Following the Polarity Theory\cite{tomkins1963left}, human worldviews are defined by humanism, which relies on humanity and experiences as intrinsically valuable, and normativism, which relies on values determined by external norms and ideals\cite{nilsson2014humanistic}. Thus, humanistic people rarely engage in misconduct because they hold group values and beliefs and seek affiliated interests\cite{singh1990managerial}. Humanism was found related positively to the political left, preferences for equality, openness, emotionality and honesty/humility traits, as well as to moral intuitions pertaining to fairness and prevention of harm, and negatively to levels of authoritarianism, social dominance, general and economic system justification\cite{nilsson2016humanistic}. By contrast, normativism is related positively to the political right, conservative issue preferences, resistance to change, acceptance of inequality, and negatively to openness, emotionality and honesty/humility traits\cite{nilsson2020rediscovering}, as well as with  moral intuitions pertaining to ingroup loyalty, respect for authority, and protection of sanctity\cite{nilsson2016humanistic}. Given this picture, it is reasonable to assume that humanism is more compatible with idealism, whereas normativism with relativism. However, normativism could be also associated with idealism if people idealize their worldview. In the present study we used only the normativism worldview scale\cite{nilsson2021death} because the humanism worldview resembles the honesty/humility personality trait.

Based on the literature reviewed above, the ethics positions (idealism and relativism), personality (honesty/humility, conscientiousness, Machiavellianism and narcissism) and worldview (normativism) variables were combined by a clustering approach in order to identify ethical profiles.

\subsection{Methods}

\subsubsection{Participants}
330 participants (182 females, 138 males and 10 individuals with no gender information; mean age = 20.52 years, s.d. = 2.55; age range = 18-35 years) were recruited from the University of L'Aquila, Italy. Participants were not requested to give their informed consent, given that the questionnaire was anonymous, in compliance with the European privacy legislation GDPR. The internal review board approved the study. 

\subsubsection{Materials}
The following tests were administered. 

\noindent
\textit{The Italian adaptation of the Ethics Position Questionnaire} (EPQ-5) \cite{o2021individual}: 5 items for idealism (e.g., A person should make certain that their actions never intentionally harm another even to a small degree 'also actions performed by Internet or the computer') and 5 for relativism (e.g., What is ethical varies from one situation and society to another 'also situations that occur on the Web'). Given the purpose of the present research, beside each item, we reported an example of an action referred to the digital world (the examples are not included in the original questionnaire). For each item, participants were asked to rate the degree of agreement from 1 (strongly disagree) to 5 (strongly agree). The item analysis showed that the corrected item-total correlation was acceptable for both scales (greater or equal to .30\cite{Nunnally&Bernestein1994}). The reliability (internal consistencies) was $\alpha$ = .63 for the idealism scale, and $\alpha$ = .69. for the relativism scale.

\noindent    
\textit{The Italian adaptation of the HEXACO-60}\cite{ashton2009hexaco}: 10 items for honesty/humility (e.g., I wouldn't use flattery to get a raise or promotion at work, even if I thought it would succeed) and 10 for conscientiousness (e.g, I plan ahead and organize things, to avoid scrambling at the last minute). For each item, participants are asked to rate the degree of agreement from 1 (strongly disagree) to 5 (strongly agree). The item analysis showed that 3 items for each scale were not satisfactory in terms of corrected item-total correlation\cite{Nunnally&Bernestein1994}. Thus, 7 items were retained for each scale. The reliability was $\alpha$ = .68 for the honesty/humility scale, and $\alpha$ = .69 for the conscientiousness scale.

\noindent  
\textit{The Italian version of the Dark Triad Dirty Dozen}\cite{schimmenti2019exploring}: 4 items for Machiavellianism (e.g. I tend to manipulate others to get my way) and 4 for narcissism (e.g. I tend to want others to admire me). For each item, participants were asked to rate the degree of agreement from 1 (strongly disagree) to 5 (strongly agree). The item analysis showed that the corrected item-total correlation was acceptable for both scales. The reliability was $\alpha$ = .77 for the Machiavellianism scale, and $\alpha$ = .81 for the Narcissism scale.

\noindent  
\textit{The English version of normativism scale} \cite{nilsson2021death}: 2 items were original (e.g. The maintenance of law and order is the most important duty of any government), while the third one was created by us. For each item, participants are asked to rate the degree of agreement from 1 (strongly disagree) to 5 (strongly agree). The item analysis showed that all items were satisfactory in terms of corrected item-total correlation. The reliability was $\alpha$ = .72.
 
\noindent      
\textit{The digital behavior scale}: 3 items created by us for privacy violation (e.g. I use the personal information of others without permission - e.g., a photo of a friend); 3 items adapted in Italian from the Unethical Computer Use Behavior Scale (UCUBS) \cite{namlu2007unethical} for copy-right infringements (e.g. I use software without owning a licence); 2 items adapted in Italian from Buchanan et al. \cite{buchanan2007development}, plus 2 created by us items for caution (e.g. Do you read a website’s privacy policy before registering your information?). For each item, participants were asked to rate the degree of frequency of each action from 1 (never) to 5 (always). The reliability was $\alpha$ = .66 for privacy violation, $\alpha$ = .66 for copy-right infringements, and $\alpha$ = .70 for caution.

\noindent    
In general, all reliability coefficients are satisfactory (see \cite{taber2018use}). 

\subsubsection{Procedure}
The questionnaire was administered by means of the platform 'LimeSurvey'. The address of the questionnaire was disseminated to the students in the the form of a URL and by a QR-Code to facilitate its use by camera-equipped devices. To avoid the order effect due to fixed order of the questions, the scales were presented randomly across participants. The questionnaire administration lasted about 15 minutes.

\subsubsection{Plan of Analysis}

Statistical analyses were performed by IBM SPSS Statistics version 20. Data were transformed in z-scores. The standard cut-off of ±3 standard deviations away from the mean was used to remove outliers: in total 13 outliers were detected and excluded from subsequent analyses. This led to a final sample of 317 participants (177 females, 130 males, and 10 individual with no gender information; mean age = 20.53 years; s.d. = 2.58). First of all, the cluster analyses were carried out using as variables idealism, relativism, honesty/humility, conscientiousness, Machiavellianism, narcissism and normativism. Two different approaches were pursued. The first approach was data-driven and aimed at exploring the dataset using the two-step method, that is conducting a hierarchical cluster analysis using Ward’s method, and a subsequent confirmatory k-means analysis. This method provides a relatively robust identification of clusters\cite{taylor2001posttraumatic}. The second approach was model-driven and aimed at confirming the ETP, which identifies four moral philosophies: situationism, subjectivism, absolutism, and exceptionism. Thus, only the k-means confirmatory analysis was used, with 4 predetermined clusters. For each approach, univariate  analyses of variance (Anova) were conducted to explore the validity, using idealism, relativism, honesty/humility, conscientiousness, Machiavellianism, narcissism and normativism as dependent variables, and the cluster solution as the independent variable. In addition, the discriminant analysis was performed to clarify the classification of the cases according to the cluster solution. Finally, Anovas were also conducted to explore the predictive power of the cluster solutions in terms of digital behavior. 

\subsection{Results}
 
\begin{figure}[!htb]
   \begin{minipage}{0.48\textwidth}
     \centering
     \includegraphics[width=1\linewidth]{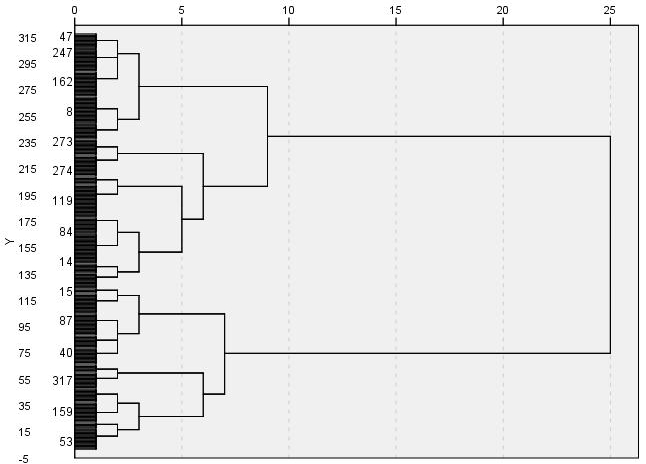}
     \caption{\raggedleft Dendrogram using the Ward's method}\label{Fig:Data1}
   \end{minipage}\hfill
   \begin{minipage}{0.48\textwidth}
     \centering
     \includegraphics[width=1.2\linewidth]{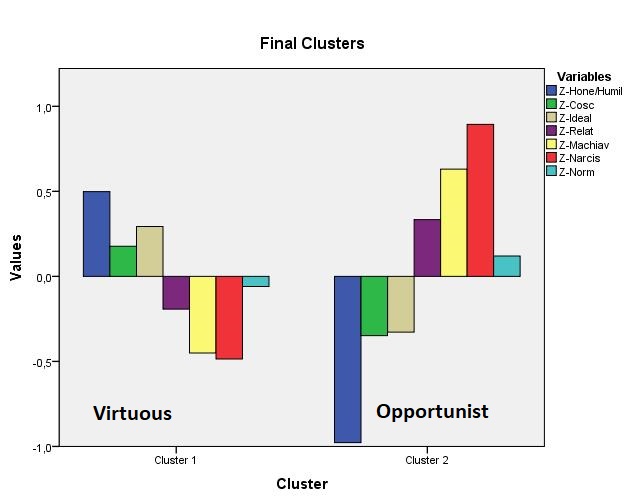}
     \caption{2-Cluster solution}\label{Fig:Data2}
   \end{minipage}
\end{figure}

\paragraph{The cluster analysis - Data-driven approach} 
The cluster solution was determined considering the dendrogram, the agglomeration schedule coefficients, and the interpretability of the cluster solution\cite{aldenderfer1984cluster}. The analysis suggested a 2-cluster solution (see Figure \ref{Fig:Data1}).

The k-means analysis was carried with $k=2$ (see Figure \ref{Fig:Data2}). The silhouette value, which is a measure of how similar an object is to its own cluster compared to the other clusters, is 0.3, which is considered acceptable. The first cluster was formed by 210 subjects, whereas the second cluster by 107 subjects. Based on the variables distribution within each cluster, we defined the first cluster 'virtuous' and the second cluster 'opportunist'.

Regarding the validity of the cluster solution, the 'virtuous' cluster produced higher scores in idealism, honesty/humility, conscientiousness, and lower scores in relativism, Machiavellianism and narcissism than the 'opportunist' cluster. No difference was found between the two clusters in normativism (see Table \ref{tab:tab2}).

Regarding the discriminant analysis, 98,7\% of original grouped cases and 98.1\% of cross-validated grouped cases were correctly classified.

\setlength{\tabcolsep}{0.35em} 
{\renewcommand{\arraystretch}{0.2}
\begin{table}[!htp]
\centering
\tiny
\begin{tabular}{ |p{2.3 cm} | p{2.5 cm}|p{3 cm}|p{3.5 cm}|}

\hline
&\textbf{ Virtuous: Mean (SD)} & \raggedright \textbf{Opportunist: Mean (SD)} &\textbf{ Statistics: Anova}\\
\hline
\textbf{Idealism}	&.293 (.055) &	-.328 (.077)	& F (1,315) = 43.25, p $<$ .0001 \\
\hline
\textbf{Relativism}	& -.193(.066) &	.333(.093)	& F (1,315) = 21.16, p $<$ .0001 \\
\hline
\textbf{Honesty/Humility} &	.498(.049) &	-.978(.069) &	F (1,315) = 300.85, p $<$ .0001\\
\hline
\textbf{Conscientiousness }&	.177(.067) &	-.348(.094) &	F (1,315) =20.82, p $<$ .0001\\
\hline
\textbf{Machiavellianism} &	-.450(.048)	& .631(.067) &	F (1,315) = 174.73, p $<$ .0001\\
\hline
\textbf{Narcissism}	& -.486(.050) &	.894(.071) &	F (1,315) = 253.66 p $<$ .0001\\
\hline
\textbf{Normativism	}& -.060(.069) &	.120(.097)	& F (1,315) = 2.290, p = .13 \\
\hline

\end{tabular}
  \caption{Differences between the ‘virtuous’ and ‘opportunist’ clusters}
    \label{tab:tab2}

\end{table}
}

As concerns the differences between the two clusters in the digital behaviour, the 'virtuous' cluster scored lower in privacy violations [F (1,315) = 26.09, p $<$ .0001; Mean(SD) -.197(.066) vs .387(.093)] and copyright infringements [F (1,315) = 29.89, p $<$ .0001; Mean(SD) -.210(.066) vs .412(.093)], and higher in caution [F (1,315) = 6.37, p $<$ .05; Mean(SD) .100(.068) vs -.197(.096)] than the 'opportunist' cluster.

\begin{figure}[!htp]
    \centering
    \includegraphics[width=0.8\linewidth]{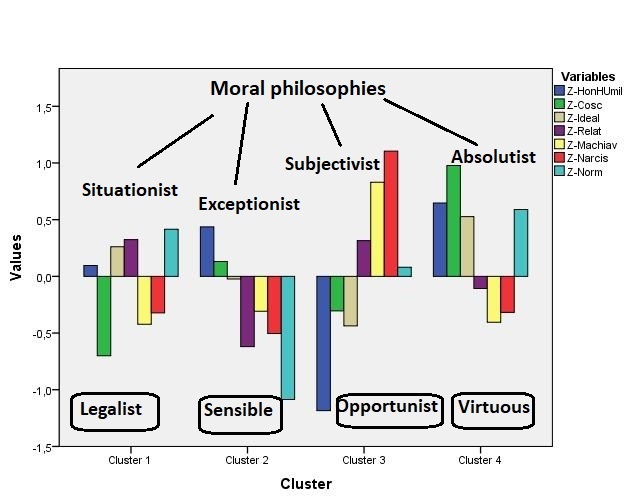}
    \caption{4-Cluster solution}
    \label{fig:cluster4}
\end{figure}

\paragraph{The cluster analysis - Model-driven approach} 
The cluster solution based on the four moral philosophies of the EPT also showed an acceptable silhouette value (.20) (see Figure \ref{fig:cluster4}).

The first cluster was formed by 86 subjects, the second by 79 subjects, the third by 77, and the fourth cluster by 75. Based on the variables distribution within each cluster, we defined the first cluster 'legalist', which related to the situationism moral philosophy, showing high scores in both idealism and relativism; the second cluster 'sensible', which related to the exceptionist moral philosophy, showing low scores in both idealism and relativism; the third cluster 'opportunist', which related to the subjectivist moral philosophy, showing a low score in idealism and a high score in relativism; the fourth cluster 'virtuous', which related to the absolutist moral philosophy, showing a high score in idealism and a low score in relativism. Regarding the validity of the cluster solution, the 'legalist', 'sensible' and 'virtuous' clusters showed no differences in terms of Machiavellianism and narcissism; the 'legalist' and 'virtuous' cluster did not differ in normativism; the 'legalist cluster did not differ in idealism with respect to the 'sensible' and 'virtuous' clusters and in relativism with respect to the 'opportunist' cluster; then, the 'sensible' and 'virtuous' cluster showed no difference in terms of honesty/humility; the other comparisons were significant (see Table \ref{tab:tab3}). 

Regarding the discriminant analysis, \%96.8 of original grouped cases and \%95.3 of cross-validated grouped cases were correctly classified. 

\setlength{\tabcolsep}{0.35em} 
{\renewcommand{\arraystretch}{0.2}

\begin{table}[!htp]
\centering
 \begin{threeparttable}
\tiny
\begin{tabular}{ |p{2.3 cm} | p{1.3 cm}|p{1.3 cm}|p{1.3 cm}|p{1.3 cm}|p{2.2 cm}|}

\hline
&\textbf{Legalist Mean (SD)} & \raggedright \textbf{Sensible Mean (SD)} &\textbf{Virtuous Mean (SD)} & \raggedright \textbf{Opportunist Mean (SD)} & \textbf{Statistics Anova} \\
\hline
\textbf{Idealism} & .261 $(.083)^a$ & -.023 $(.087)^b$ 	& .527 $(.089)^{bc}$  &	-.437 $(.088)^{abc}$ & 	F (3,313) = 21.81, p $<$ .0001 \\
\hline
\textbf{Relativism}	& .324 $(.099)^a$ &  	-.619 $(.103)^{ab}$ & -.106 $(.106)^{abc}$ & .315 $(.105)^{bc}$ &	F (3,313) = 18.87, p $<$ .0001 \\
\hline
\textbf{Honesty/Humility} &	.095 $(.077)^a$ & 	.437 $(.081)^{ab}$ &	.647 $(.083)^{ac}$  &	-.1184 $(.082)^{abc}$ &	F (3,313) = 100.75, p $<$ .0001\\
\hline
\textbf{Conscientiousness } & -.700 $(.085)^a$	& .131 $(.088)^{ab}$ & .978 $(.091)^{abc}$ & -.304 $(.090)^{abc}$ &	F (3,313) = 65.91, p $<$ .0001\\
\hline
\textbf{Machiavellianism} & -.423 $(.074)^a$ &	-.308 $(.077)^b$ &	-.405 $(.079)^c$ & .830 $(.078)^{abc}$ &	F (3,313) = 61.28, p $<$ .0001\\
\hline
\textbf{Narcissism}	& -.322 $(.080)^a$ & -.504 $(.083) ^b$ & -.318 $(.086)^c$ &	1.105 $(.085)^{abc}$ &	F (3,313) = 79.05, p $<$ .0001\\
\hline
\textbf{Normativism	}& .415 $(.082)^a$ &	-1.086 $(.086)^{ab}$ & 	.589 $(.088)^{bc}$ & .080 $(.087)^{abc}$ &	F (3,313) = 76.78, p $<$ .0001 \\
\hline

\end{tabular}
\begin{tablenotes}
      \tiny
  \item[*] Significant differences between clusters are showed by letter correspondences; Tukey’s Post-hoc (HSD) were used; $^a$ Legalist; $^b$ Sensible; $^c$ Virtuous
    \end{tablenotes}
  \caption{Differences between the ‘legalist’, ‘sensible’, ‘virtuous’ and ‘opportunist’ clusters}
    \label{tab:tab3}
 \end{threeparttable}
\end{table}
}

As concerns the differences between the four clusters in the digital behaviour, the 'legalist', 'virtuous' and 'sensible' clusters showed no difference in privacy violation; the 'opportunist' cluster showed higher scores than the other 3 clusters. Regarding copy-right infringements, the 'legalist' cluster showed no difference as compared to the 'sensible' and 'virtuous' clusters; the 'sensible' cluster showed higher scores than the 'virtuous' cluster; then, the 'opportunist' cluster showed higher scores than the other 3 clusters. Regarding caution, the 'virtuous' cluster showed higher scores only than the 'opportunist' cluster; the other comparisons were not significant (see Table \ref{tab:tab4}). 

\setlength{\tabcolsep}{0.35em} 
{\renewcommand{\arraystretch}{0.2}

\begin{table}[!htp]
\centering
 \begin{threeparttable}
\tiny
\begin{tabular}{ |p{1.7 cm} | p{1.3 cm}|p{1.3 cm}|p{1.3 cm}|p{1.3 cm}|p{2 cm}|}

\hline
&\textbf{Legalist Mean (SD)} & \raggedright \textbf{Sensible Mean (SD)} &\textbf{Virtuous Mean (SD)} & \raggedright \textbf{Opportunist Mean (SD)} & \textbf{Statistics Anova} \\
\hline
\textbf{Privacy violation} & -.065 (.104)$^a$	& -.150 (.109)$^b$ &	-.246 (.112)$^c$ &	.466 (.110)$^{abc}$ &	F (3,313) = 8.35, p $<$ .0001 \\
\hline
\raggedright \textbf{Copy-right infringements} & -.097 (.104)$^a$ &	.034 (.108)$^b$ &	-.374 (.111)$^{bc}$ &	.438 (.110)$^{abc}$ &	F (3,313) = 9.41, p $<$ .0001 \\
\hline
\raggedright \textbf{Caution} & .029 (.106)	 & .048 (.111)	& .226 (.114)$^a$ &	-.302 (.112)$^a$ &	[F (3,313) = 3.80, p $<$ .05 \\ 
\hline

\end{tabular}
\begin{tablenotes}
      \tiny
  \item[*] Significant differences between clusters are showed by letter correspondences; Tukey’s Post-hoc (HSD) were used; $^a$ Legalist; $^b$ Sensible; $^c$ Virtuous
    \end{tablenotes}
  \caption{Differences between the ‘legalist’, ‘sensible’, ‘virtuous’ and ‘opportunist’ clusters}
    \label{tab:tab4}
 \end{threeparttable}
\end{table}
}


%% file: Discussion.tex
The present study was aimed at creating ethical profiles that can be predictive of digital behaviors. Ethical profiles were created by two different clustering approaches, data-driven and model-driven combining the Ethics Position Theory (idealism and relativism) with personality (honesty/humility, conscientiousness, Machiavellianism, and narcissism) and worldview (normativism) variables. 

The analyses showed that the 2-cluster solution is actionable because it shows an overlapping between the clusters only in terms of normativism. In addition, the discriminant analysis showed that the original grouped cases were correctly classified with a large percentage. As expected, the 'virtuous' cluster tends to commit less privacy violation and copyright infringements than the 'opportunist' cluster. This suggests that the 'virtuous' cluster is guided by principled values with respect to their counterpart, which instead is more prone to violate rules in order to achieve personal benefits. This result extends to digital behaviors previous findings concerning the ethical profiles in marketing negotiations\cite{al2016ethical}. In addition, the 'virtuous' cluster scored higher in privacy setting and information registration on internet (caution) than the 'opportunist' cluster. This suggests that the 'virtuous' cluster pay more attention, and are more cautious than the 'opportunist' cluster. 

As concerns the 4-cluster solution, although the original grouped cases were correctly classified with a large percentage, the results showed some overlap between the 'legalist', 'virtuous' and 'sensible' clusters, especially in terms of Machiavellianism and narcissism. Only the 'opportunist' cluster showed higher scores than the other 3 clusters in privacy violation and copy-right infringements. The 'legalist' cluster showed lower scores only than the 'opportunist' cluster in terms of copyright infringements, suggesting that legalists pay attention to this matter as well as sensible and virtuous people. Yet, only the 'virtuous' cluster showed lower scores than the 'opportunist' cluster in caution. In general, these results show that the 'legalist', 'virtuous' and 'sensible' clusters are characterized basically by positive attributes. 

Taken together, these findings suggest that the 2-cluster solution can represent a starting point for refinement ethical profiles. The 4-cluster solution is rather a work in progress, although it partially confirmed the EPT. It should be validated by differentiating better the clusters. 

%% file: Conclusions_and_Future_Work.tex

These findings are very promising and open many directions for refinements. Future works are aimed at better disentangling not only the 'legalist', 'sensible' and 'virtuous' clusters, but also the 'opportunist' cluster in sub-clusters. This might be achieved by considering other variables. For example, one can include other scales measuring integrity, autenticity, dogmatism, moral disengagement, sadism, opportunism, normlessness, and so forth. In terms of personality variables one can add neuroticism and openness to experience. Referring to the 'virtuous' and 'opportunist' clusters, it would be interesting to clarify if they can be unpacked in more specific clusters. Indeed, we used mainly moral variables and morally oriented personality traits. Variables that are more oriented to measure normative and consequential ethics, such as legalism, rationality, utilitarianism, authority, and so forth, could also be explored. In addition, we intend to investigate if the notion of ”semantic clustering”, for which the similarity of different concepts in a taxonomy is measured by computing their edge distance within the taxonomy \cite{althobaiti2017comparison}, could help
when building user’s profile. In order to do so, we are working 
to elaborate a taxonomy out of the questionnaire data, i.e. a 
structure to which a semantic distance can be applied. 
Having the data organized in a 
structured
way could also be useful for applying different 
clustering techniques (such as Graph Neural Network\cite{4700287}). Finally, the findings provided by this study could be also generalized to other digital behaviors, such as those related to social networks.